\newif\ifanonymousmode
\anonymousmodefalse%

\ifanonymousmode
  \PassOptionsToClass{anonymous}{acmart} %
\fi

\documentclass[sigconf]{acmart} %

\renewcommand{\footnotetextcopyrightpermission}[1]{}
\usepackage{hyperref}
\usepackage{graphicx}
\title{Quantum est in Libris: Navigating Archives with GenAI, Uncovering Tension Between Preservation and Innovation}

\author{Mar Canet Sola}
\orcid{0000-0001-5986-3239}
\affiliation{%
  \institution{BFM, Tallinn University, Estonia}
  \city{}
  \country{}}
\affiliation{%
  \institution{Academy of Media Art Cologne(KHM), Germany}
  \city{}
  \country{}}
\email{mar.canet@gmail.com}

\author{Varvara Guljajeva }
\authornote{Both authors contributed equally}

\orcid{0000-0002-0261-3121}
\affiliation{%
  \institution{ VCUarts Qatar, Qatar}
  \country{}
}
\affiliation{%
  \institution{ Academy of Media Art Cologne, Germany}
  \country{}
}
\email{varvarag@gmail.com}

\ifanonymousmode
\else
  \renewcommand{\shortauthors}{Mar Canet Sola, Varvara Guljajeva}
\fi

\begin{document}

\acmConference{}{}{}
\acmBooktitle{}
\acmYear{}
\copyrightyear{}
\acmDOI{}
\acmISBN{}
\acmPrice{}

\begin{abstract}

"Quantum est in libris" explores the intersection of the archaic and the modern. On one side, there are manuscript materials from the Estonian National Museum's (ERM) more than century-old archive describing the life experiences of Estonian people; on the other side, there is technology that transforms these materials into a dynamic and interactive experience. Connecting technology and cultural heritage is the visitor, who turns texts into inputs for a screen sculpture.

Historical narratives are visually brought to life through the contemporary technological language. Because the video AI models we employed, Runway Gen-3 and Gen-4, have not previously interacted with Estonian heritage, we can observe how machines today "read the world" and create future heritage. "Quantum est in libris" introduces an exciting yet unsettling new dimension to the concept of cultural heritage: in a world where data are fluid and interpretations unstable, heritage status becomes fragile. In the digital environment, heritage issues are no longer just about preservation and transmission, but also about representation of the media, machine creativity, and interpretive error. Who or what shapes memory processes and memory spaces, and how?

\end{abstract}



\begin{teaserfigure}
    \centering
    \includegraphics[width=1\linewidth]{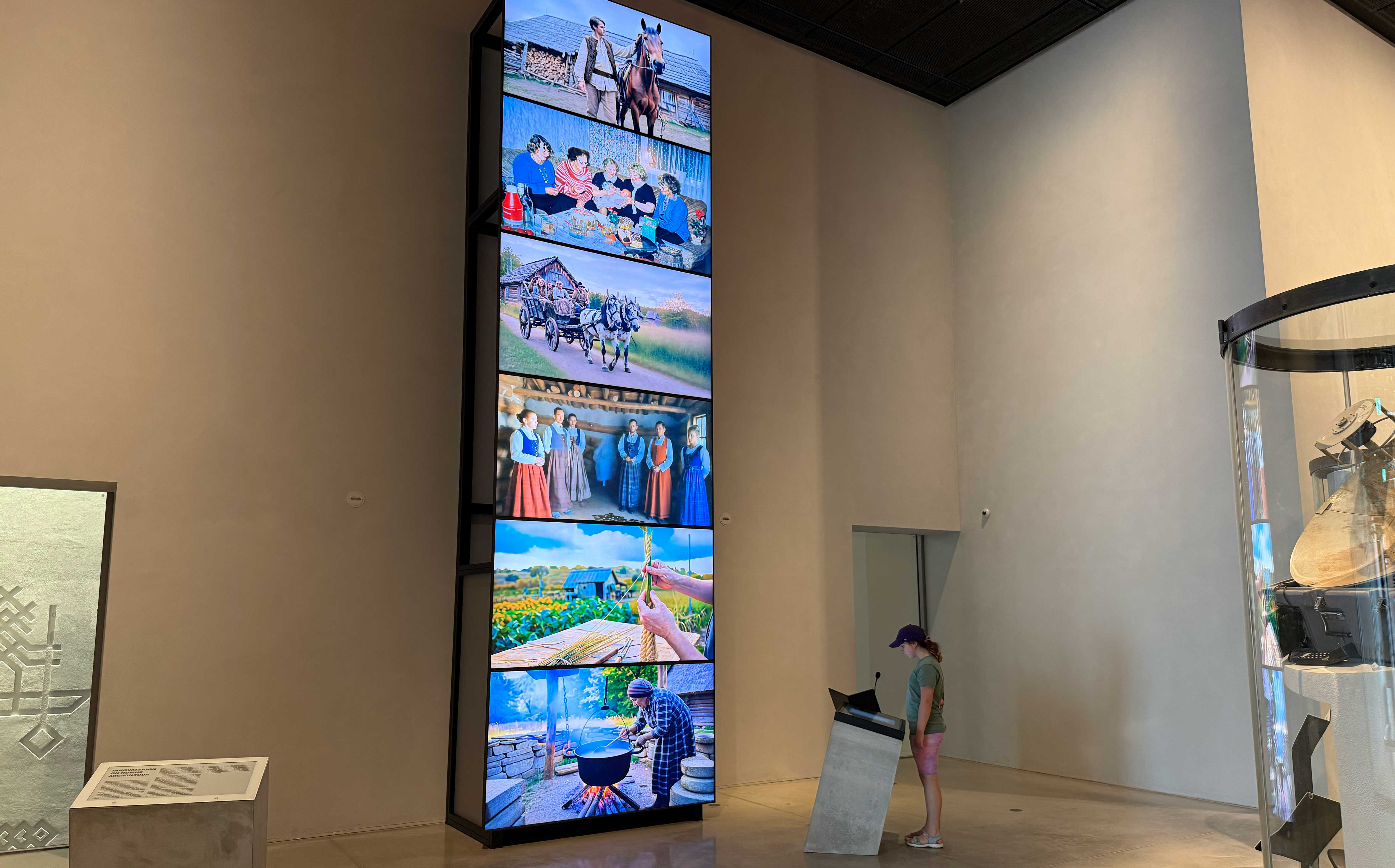}
    \caption{Installation view of Quantum est in Libris in the museum}
    \label{fig:enter-label}
\end{teaserfigure}

\maketitle

\section{INTRODUCTION}
\section{Introduction}

In 2025, Estonia marks the 500th anniversary of the earliest known Estonian-language book, underscoring the enduring relationship between textual heritage and national identity such as singing~\cite{ smidchens2014singing}. Over centuries, the technologies of writing, reading, and preserving collective memory have evolved from manuscripts and printed volumes to the digital archives of today~\cite{deegan2006digital}. Contemporary museums and archives now face the challenge and opportunity not only of protecting artifacts but also of reimagining public engagement with heritage through emerging technologies~\cite{giannini2019artificial, giaccardi2012heritage}.

The intersection of cultural heritage and artificial intelligence (AI) has rapidly become a focus of critical scholarship~\cite{giannini2019artificial}. AI techniques, particularly those that exploit large-scale generative models, are opening new avenues to interpret, visualize, and interact with historical materials~\cite{miller2022artificial}. Such technologies can augment curatorial work, enable creative reinterpretations, and bring new audiences into conversation with the past. However, these same systems often have biases that reflect their training data, leading to gaps in cultural representation, especially for smaller languages and communities~\cite{bender2021dangers}. This tension between technological innovation and the preservation of authentic memory is at the heart of contemporary debates in digital heritage and media art~\cite{giaccardi2012heritage}.

Our own work has engaged with the idea of \emph{future heritage} since 2012, beginning with the \emph{NeuroKnitting} project, which augmented traditional craft with EEG data by translating brainwave patterns into textile structures~\cite{guljajeva2012neuroknitting}. In 2022, \emph{NeuroKnitting Beethoven} extended this approach by mapping the brainwaves of a pianist interpreting Beethoven into knitting machine processes~\cite{guljajeva2023interactive}. Beyond their innovative practice, both projects re-engaged with cultural heritage while simultaneously creating new forms of heritage—patterns shaped by neural activity that may hold future cultural significance~\cite{guljajeva2023telematic}. In a similar way, the current project \emph{Quantum est in Libris} explores how current tools, such as generative AI, can enter into dialogue with past cultural heritage to create new modes of future heritage for generations to come.

Engaging with these debates requires considering heritage not only as preservation of the past but also as projection into possible futures. Future heritage can be understood as a testing ground for imagining what future generations might value and wish to experience. As Holtorf and Högberg emphasize in their book \emph{Contemporary Heritage and the Future}, "Heritage management is a futuristic activity because to a large extent it is motivated by the present-day desire to preserve the remains of the past for the benefit of future generations"~\cite{holtorf2015contemporary}.

\section{ARTWORK DESCRIPTION}
This interactive artwork invites participants to explore a curated selection of 84 historical handwritten pages from the ethnographical diaries. This archival material, carefully chosen by the research personnel of ERM and the artists, offers insight into Estonian life in earlier times. This piece creates a new method for public engagement with heritage materials such as old diaries using AI videos and interactivity, bringing archival material with significant historical and cultural value into an engaging, interactive experience for the public.

\begin{figure*}[t]
    \centering
    \includegraphics[width=1\textwidth]{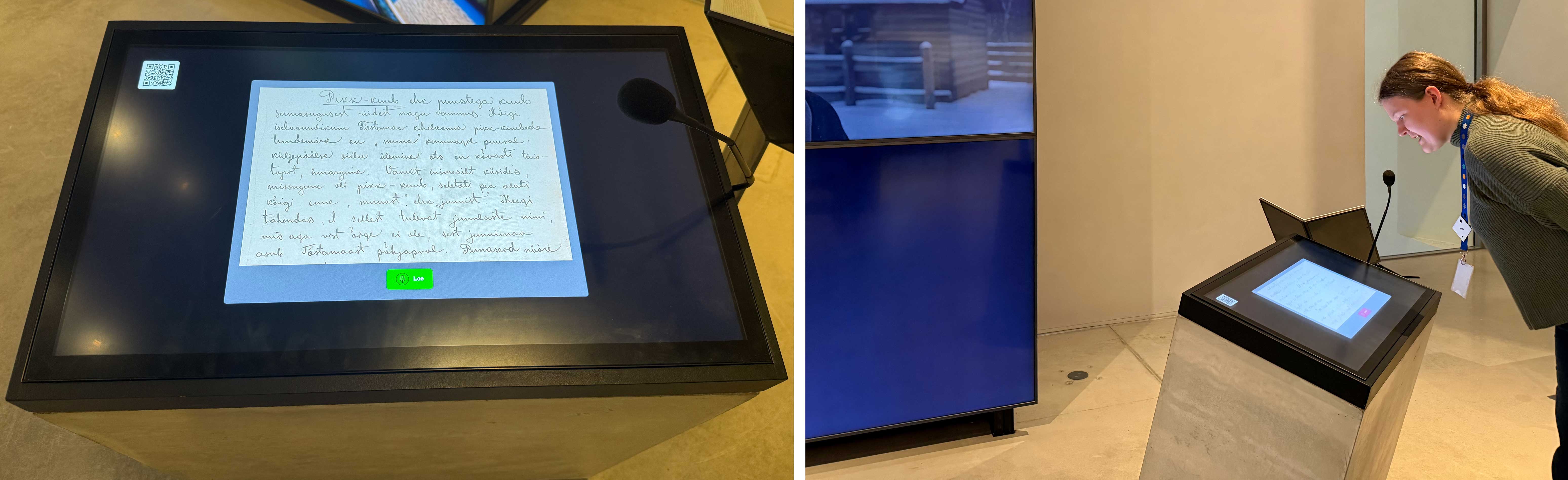}
    \caption{Top: the interface. Bottom: a participant interacting with the art piece by reading a page from the diary}
    \label{fig:speaking}
\end{figure*}

\section{ABOUT THE ARTWORK}
\subsection{PROJECT CONCEPT}

In 2025, the 500th anniversary of the written Estonian marks the 500th anniversary of the writing. Over centuries, the ways we write, read, work, and live have changed significantly. Technology is evolving rapidly and offers new possibilities for the preservation and new artistic innovative ways to present culture such as archive materials.

"Quantum est in Libris" explores the meeting point between the archaic and the contemporary. On the one hand are handwritten materials from the archive of the Estonian National Museum, more than a century old, describing the life experiences of Estonians. The other side is technology, turning these materials into a dynamic and interactive experience.

At the heart of the installation is a sculpture consisting of a totem of six screens along with an interactive station. Visitors read aloud randomly selected excerpts from historical texts, which the installation records and transforms into visual imagery traveling across the screen sculpture. The images displayed on the screens are generated based on interactions with the five most recent visitors. The installation bridges past and present, transforming centuries-old descriptions of daily life into digital art through visitor engagement.

"Quantum est in Libris" reflects how contemporary machines perceive the world and how humans navigate information in the age of artificial intelligence. It is a visual spectacle in which historical narratives come alive through the language of contemporary technology.

\begin{figure}[H] 
    \centering
    \includegraphics[width=1\linewidth]{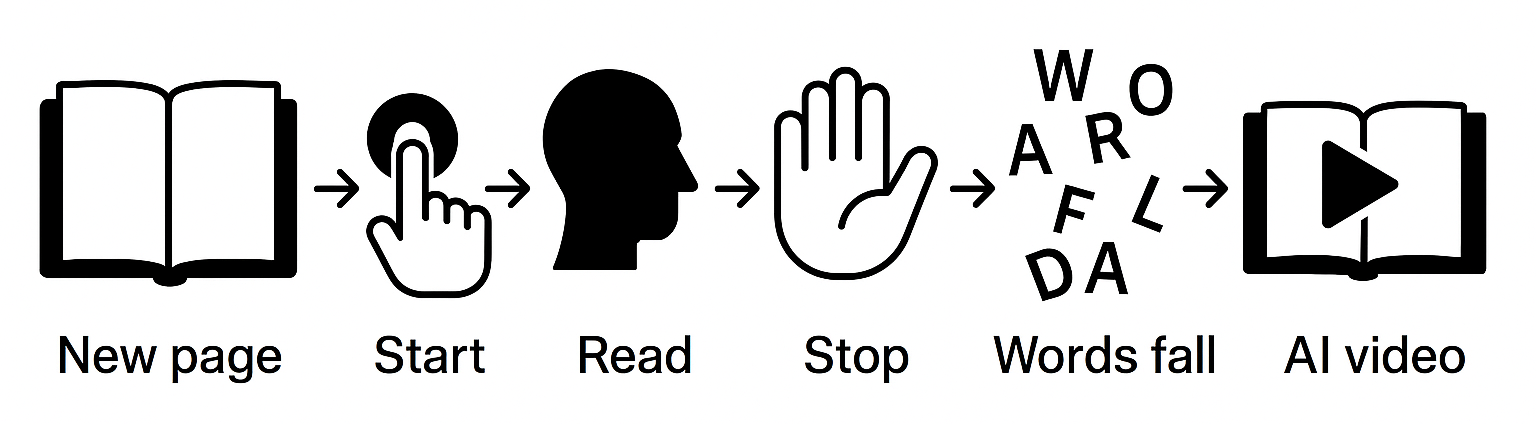}  
    \caption{Diagram of the steps of the interaction flow}
    \label{fig:interaction-flow}
\end{figure}

\subsection{WORKFLOW}

This project follows this workflow:

\begin{enumerate}
    \item New page from diaries of ERM Museum Archive is selected and display.
    \item The participant presses a button to begin reading.
    \item The participant reads the page aloud.
    \item The participant stops the reading.
    \item The words spoken fall onto the screen.
    \item Once all words have fallen, they enlarge and dissolve, revealing a new AI-generated video emerging from the diary page in the bottom screen.
\end{enumerate}

\begin{figure*}[t]
 
    \centering
    \includegraphics[width=1\textwidth]{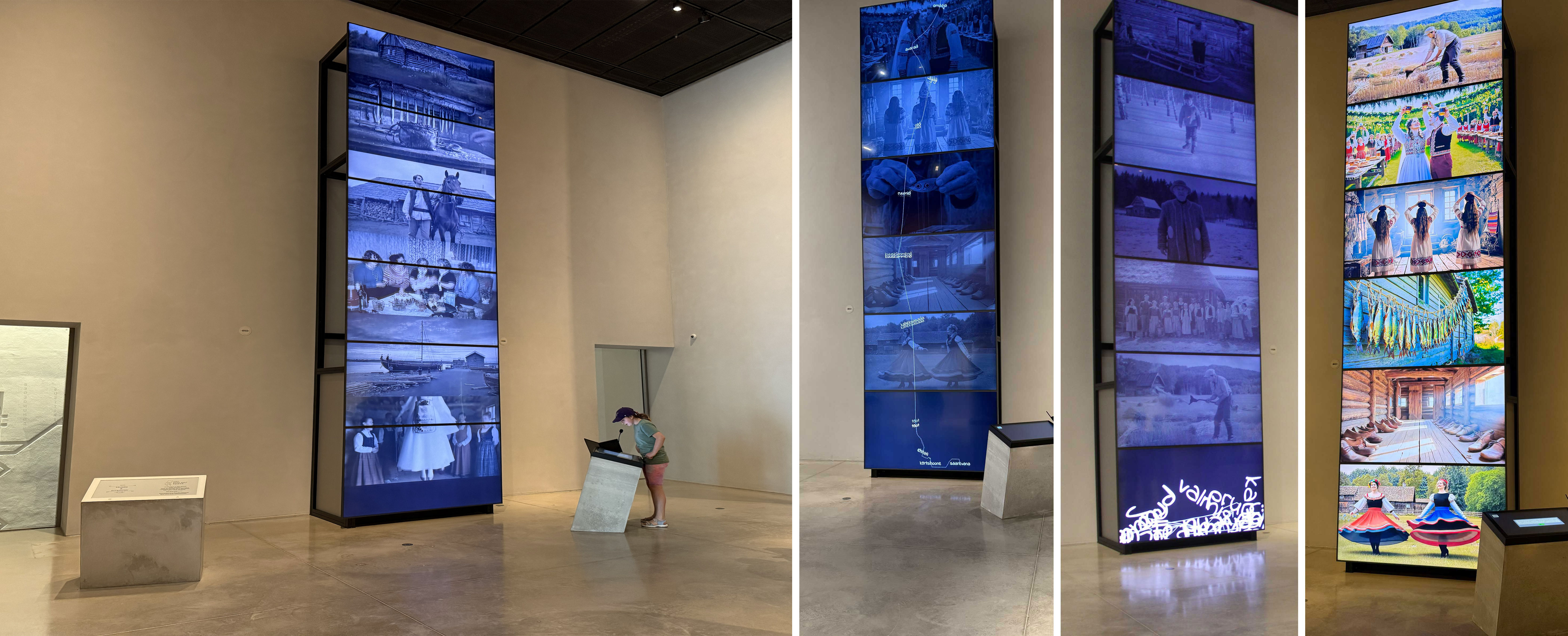}
    \caption{Four states of the installation (left to right): (a) Start reading (b) falling words triggered by interaction; (c) enlarged words followed by an AI-generated video of the diary page at the base of the screen tower; (d) installation view showing the last six audience contributions, i.e., videos generated from archival texts read by the public.}
    \label{fig:different-screens}
\end{figure*}

The Figure \ref{fig:interaction-flow} shows an overview of the interaction flow. In Figure \ref{fig:speaking} illustrates steps 1-3. The rest of the steps are illustrated in Figure \ref{fig:different-screens}. After 1 minute, the system, having moved to step 6, starts again at step 1, waiting for new interaction. Meanwhile, the screen tower shows six videos in a perfectly seamless loop, where the transition between the last and first frame is imperceptible, with the newer ones at the bottom and the older ones at the top.

\subsection{TECHNICAL DESCRIPTION}

The installation is operated by a system of two computers and three main applications, each with a specific role.

\paragraph{Display Computer}
This computer with an openFrameworks C++ app renders all visual content, including AI videos and text animations, which are displayed on six synchronized full-HD screens.

\paragraph{Interaction Station}
This second computer manages all user interaction. It is connected to a touch screen and a microphone and runs two applications:
\begin{itemize}
    \item \textbf{Backend Server:} A custom application that manages all internal communication between the components of the three apps and serves the web-based front-end.
    \item \textbf{Front-end Interface:} A web application running in Google Chrome's kiosk mode. This interface is displayed on the touch screen, allowing participants to interact with the exhibit. The speech-to-text rely on the Web Speech API in Estonian and English.
\end{itemize}

\paragraph{Web Interface and Archive Link}
The user interface on the touch screen displays digitized diary pages. Alongside each page is a QR code that links directly to the object's official entry in ERM archive, hosted on MUIS\footnote{\url{https://www.muis.ee/}} (the Estonian National Archive Repository). This feature allows participants to explore the original source material and discover additional contextual information.

\subsection{From Archives to Moving Images}

The production of the AI-generated videos was the most time-consuming aspect of the project. Although recent tools have made generating videos with AI easier and faster, producing longer sequences remains a labor-intensive process clips beyond ten seconds still demands significant time and effort. Each video in the installation attempts to visually represent the activity described in the handwritten text of a single diary page and is generated as a seamless, perfectly looping sequence. Achieving such loops is a distinctive feature of AI-generated content and is almost impossible with traditional camera-based video production. The durations of the videos vary between 20 seconds and 2 minutes 40 seconds. Each required multiple generations and careful selection to best capture the intended ideas. Our previous work with StyleGAN involved creating custom datasets, which took weeks to produce approximately 20 minutes of video material~\cite{guljajeva2022postcard}. Later, we adopted Stable Diffusion for real-time visual generation, which streamlined certain aspects of experimentation~\cite{canet2024visions}. For this project, we turned to Runway GEN3 and the more recent GEN4 models. Notably, the outputs of these two models differ in resolution, complicating their integration. For longer videos, several clips had to be joined together, since GEN3 allowed extensions up to 40 seconds, while GEN4 only generated fixed clips of 5 or 10 seconds without extensions. This made GEN3 more suitable for our creative task, although we still employed GEN4 for selected videos.

Although the videos are generated by AI, considerable human labor is involved in their creation. The composition and, especially, the selection from the many generated iterations were performed manually. One major difficulty encountered was the limited availability of Estonian and Baltic-specific material in AI models’ training data, as Estonia is a relatively small country compared to others. This challenge required us to refine our prompts, often specifying in detail the objects and actions needed, and to conduct many generations to find a satisfactory representation. Because this process was arduous and the text-to-video models remain in development, we decided to intentionally retain certain errors in the final videos. These errors—such as, for example, a figure walking backwards—clearly indicate that the content was generated by AI, and contribute to a unique aesthetic. Moreover, since the diary texts are simple descriptions of actions and ways of living in historical Estonia, the resulting videos function as imperfect translations, in the sense described by semiotician Juri Lotman, for whom the translation between different semiotic spaces is a source of creativity~\cite{ibrus2020creativity}.

Initially, we considered building a real-time pipeline for video generation. However, this approach proved problematic because, when representing Estonian heritage, the AI-generated results would frequently fail to be culturally accurate or meaningful. The prompts often do not directly correspond to the source text and require multiple attempts to yield outputs suitable for curation by a human expert. Thus, a fully autonomous system was not feasible for this project at the current state of technology; in fact, a human-curated approach is better suited to the complexities of working with heritage materials.

\section{DISCUSSION}

\textit{Quantum est in Libris} responds to this context by bringing together handwritten archival materials from the ERM and cutting-edge text-to-video AI models—specifically, Runway Gen-3 and Gen-4\footnote{https://research.runwayml.com/}. In the installation, museum visitors read excerpts from historic diaries aloud. Their speech is transcribed and transformed into AI-generated video that visually interprets the archival content. This process foregrounds both the creative agency of the participant and the interpretive mediation of the machine, situating the visitor as an active link between centuries-old documents and digital reanimation.

This type of mediation is not without its complexities. AI models rarely possess deep knowledge of localized or minority cultures; their representations of Estonian heritage, for example, may contain inaccuracies or unexpected associations. As a result, the installation exposes both the potential and the limitations of generative AI in cultural contexts. Such errors or "creative translations" echo Lotman's theory of cultural semiotics, wherein the act of translating information from one sign system (such as text) to another (such as visual media) is inherently generative and may result in new, sometimes unforeseen meanings~\cite{lotman2013unpredictable}.

In our previous research, we have examined how artists deploy generative AI as both a creative tool and a critical lens~\cite{guljajeva2023artistic}~\cite{guljajeva2024artist}. \textit{Quantum est in Libris} extends this trajectory, serving as another case study that applies generative technologies within interactive scenarios while exploring and visualizing the national archive. In doing so, it highlights how artistic practice can simultaneously critique, expand, and reframe the role of AI in cultural heritage contexts.

Beyond technical considerations, the work raises broader questions about memory, authorship, and authenticity. In digital environments, the notion of heritage becomes fluid: archives are not static repositories but sites of ongoing negotiation between humans and machines~\cite{erll2011memory}. While generative AI can expand access and foster new forms of creativity, scholars caution that it should not replace the interpretive expertise of human curators, art historians, and artists~\cite{giaccardi2012heritage, giannini2019artificial}. Instead, installations such as \textit{Quantum est in Libris} demonstrate the need to position AI as a collaborator - one that highlights the evolving dialogue between cultural heritage, technological mediation, and future imaginaries.

\section{CONCLUSION}

This paper situates \textit{Quantum est in Libris} as a case study for navigating the tensions and opportunities at the intersection of cultural heritage and generative AI. The installation demonstrates both the creative potential and the limitations of algorithmic systems in heritage contexts, particularly when engaging with underrepresented languages and communities. By staging a dialogue between archive, machine, and audience, the project not only reinterprets existing cultural memory but also contributes to the making of \emph{future heritage}. In doing so, it invites us to reconsider what preservation, innovation, and participation might mean in an era where cultural heritage is increasingly shaped through generative technologies and interactive practices.

\ifanonymousmode
\else
  \section*{ACKNOWLEDGMENTS}
    We would like to express our gratitude to ERM for their multifaceted support with the production of the art installation Quantum est in Libris.
\fi

\bibliography{references}

\end{document}